\documentclass[11pt]{article}
\usepackage{cite}
\usepackage{epsfig,subfigure}
\usepackage{amssymb, amsmath}
\usepackage{color}
\usepackage{bm}
\usepackage[normalem]{ulem}
\usepackage{mathtools}

\def\QED{\mbox{\rule[0pt]{1.5ex}{1.5ex}}}
\def\proof{\noindent\hspace{2em}{\it Proof: }}

\usepackage{graphicx}
\usepackage{color}

\usepackage{amssymb}
\usepackage{amsmath,amsfonts,amssymb}
\usepackage{verbatim}
\usepackage{stfloats}
\usepackage[bookmarks=false]{}
\usepackage{mathtools}
\usepackage{multirow}
\usepackage{verbatimbox}
\DeclarePairedDelimiter\ceil{\lceil}{\rceil}

\newtheorem{theorem}{Theorem}

\newtheorem{lemma}{Lemma}

\newcommand\blfootnote[1]{%
  \begingroup
  \renewcommand\thefootnote{}\footnote{#1}%
  \addtocounter{footnote}{-1}%
  \endgroup
}

\begin{document}
\date{}
\title{The Capacity of $T$-Private Information Retrieval \\ with Private Side Information}
\author{\small Zhen Chen, Zhiying Wang, Syed Jafar\\
{\small Center for Pervasive Communications and Computing (CPCC)}\\
{\small University of California Irvine, Irvine, CA 92697}\\
{\it \small Email: \{zhenc4, zhiying, syed\}@uci.edu}}
\maketitle

\blfootnote{
}
\begin{abstract}
We consider the problem of $T$-Private Information Retrieval with private side information (TPIR-PSI). In this problem, $N$ replicated databases store $K$ independent messages, and a user, equipped with a local cache that holds $M$ messages as side information, wishes to retrieve one of the other $K-M$ messages.  The desired message index and the side information must remain jointly private even if any $T$ of the $N$ databases collude. We show that the capacity of TPIR-PSI  is $\left(1+\frac{T}{N}+\cdots+\left(\frac{T}{N}\right)^{K-M-1}\right)^{-1}$. As a special case obtained by setting $T=1$, this result settles the capacity of PIR-PSI, an open problem previously noted by Kadhe et al. We also consider the problem of symmetric-TPIR with private side information (STPIR-PSI), where the answers from all $N$ databases reveal no information about any other message besides the desired message. We show that the capacity of STPIR-PSI is $1-\frac{T}{N}$ if the databases have access to common randomness (not available to the user) that is independent of the messages, in an amount that is at least $\frac{T}{N-T}$ bits per desired message bit. Otherwise, the capacity of STPIR-PSI is zero.
\end{abstract}

\section{Introduction}
The private information retrieval (PIR) problem  investigates the privacy of the contents downloaded from public databases. In the classical form of PIR \cite{PIRfirstjournal}, a user wishes to, as efficiently as possible, retrieve one of $K$ messages that are replicated across $N$ non-colluding databases while preserving the privacy of the desired message index. Since its first formulation by Chor et al. in  \cite{PIRfirstjournal}, the PIR problem has been studied extensively in computer science and cryptography  under both information-theoretic and computational privacy constraints \cite{Yekhanin, Beimel_Ishai_Kushilevitz, William, CPIR, SymPIR}. 
While  studies of PIR typically seek to optimize both the upload and download costs, recently there has been a  burst of activity aimed at \emph{capacity} characterizations for information-theoretic PIR under the assumption of large message sizes, so that the communication cost is dominated by the download cost \cite{Shah_Rashmi_Kannan, Sun_Jafar_BIAPIR, Sun_Jafar_PIR, Tajeddine_Rouayheb, Chan_Ho_Yamamoto, Banawan_Ulukus}. The  capacity of PIR was  defined  in \cite{Sun_Jafar_PIR} as the maximum number of bits of desired message that can be privately downloaded per bit of total downloaded information from all the servers. In order to summarize some of the capacity results for PIR, let us define the function 
$\Psi(A,B)=\left(1+A+A^2+\cdots+A^{B-1}\right)^{-1}$
for positive real number $A$ and positive integer $B$. 
Correspondingly, $\Psi(A,\infty)=1-A$ for $A<1$. The capacity of PIR was  characterized in \cite{Sun_Jafar_PIR} as $C_{\text{\tiny PIR}} = \Psi(1/N,K)$. The capacity of $T$-PIR, where the privacy of the user's desired message index must be protected against collusion among any set of up to $T$ servers, was characterized in \cite{Sun_Jafar_TPIR} as $C_{\text{\tiny TPIR}} = \Psi(T/N,K)$. The capacity of symmetric PIR (SPIR), where the user  learns nothing about the database besides his desired message, was shown in \cite{Sun_Jafar_SPIR} to be $C_{\text{\tiny SPIR}}=\Psi(1/N,\infty)$, and the capacity of STPIR, with both symmetric privacy and robustness against  collusion among any $T$ servers, was characterized in \cite{Wang_Skoglund} as $C_{\text{\tiny STPIR}}=\Psi(T/N,\infty)$. Capacity characterizations have also been found for a number of other variants of PIR, such as PIR with MDS coded storage \cite{Banawan_Ulukus}, multi-message PIR \cite{Banawan_Ulukus_MPIR}, multi-round PIR \cite{Sun_Jafar_MPIR}, secure PIR \cite{Jia_Sun_Jafar_XSTPIR}, and PIR with side information \cite{Tandon_CPIR,Kadhe_Garcia_Heidarzadeh_Rouayheb_Sprintson,Wei_Banawan_Ulukus,Wei_Banawan_Ulukus_Side, Rouayheb_Sprintson_single_coded, Rouayheb_Sprintson_single_multi, Rouayheb_Sprintson_single_multi_individual, Rouayheb_Sprintson_single_single_online, Li_Gastpar, Li_Gastpar_muPIR}. Especially relevant to this work is the problem of PIR with side information. 

The recent focus on capacity of PIR with side-information started with the work on cache-aided PIR by Tandon \cite{Tandon_CPIR}, where the user has enough local cache memory to store a fraction $r$ of all messages as side information. In this model, the side information can be any function of the $K$ messages (subject to the storage constraint) and is globally known to both the user and all the databases. The capacity for this setting is characterized in \cite{Tandon_CPIR} as $\Psi(1/N,K)/(1-r)$. 

Formulations where databases are not aware of the side information were introduced by Kadhe et al. in  \cite{Kadhe_Garcia_Heidarzadeh_Rouayheb_Sprintson}.  In their model the side information is comprised of $M$ messages cached by the user, and the databases are not aware of the identities of these $M$ cached messages. This model gives rise to two interesting variants of PIR with side information depending on whether or not the privacy of the side information must also be preserved from the databases. PIR with non-private side information (PIR-NSI) refers to the assumption that the privacy of side information need not be preserved, while PIR with private side information (PIR-PSI)  implies that the \emph{joint} privacy of the desired message and the side information must be preserved. 

For PIR-NSI with a single database $(N=1)$, Kadhe et al. showed in  \cite{Kadhe_Garcia_Heidarzadeh_Rouayheb_Sprintson} that the capacity is $\ceil{\frac{K}{M+1}}^{-1}$. The single-database setting  has seen rapid progress in various directions \cite{Rouayheb_Sprintson_single_coded, Rouayheb_Sprintson_single_multi, Rouayheb_Sprintson_single_multi_individual, Rouayheb_Sprintson_single_single_online,  Li_Gastpar_muPIR}. However, PIR-NSI with \emph{multiple} databases turns out to be considerably more challenging. In \cite{Kadhe_Garcia_Heidarzadeh_Rouayheb_Sprintson},  Kadhe et al. provide an achievable scheme for PIR-NSI with multiple databases $(N>1)$, which achieves the rate $\Psi(1/N, K/(M+1))$ when $(M+1)\mid K$.  In spite of some progress in this direction \cite{Li_Gastpar, Wei_Banawan_Ulukus_Side, Wei_Banawan_Ulukus}, the capacity of PIR-NSI generally remains open\footnote{The converse in \cite{Li_Gastpar} does not  cover the scope of PIR-NSI, because  the privacy condition assumed in   \cite{Li_Gastpar} is not a necessary condition for PIR-NSI schemes.} for multiple databases. 

For PIR-PSI with a single database $(N=1)$, Kadhe et al. found in \cite{Kadhe_Garcia_Heidarzadeh_Rouayheb_Sprintson} that the capacity is $(K-M)^{-1}$. The capacity of PIR-PSI with more than one database was  left as an open problem in \cite{Kadhe_Garcia_Heidarzadeh_Rouayheb_Sprintson}. Remarkably, neither a general achievable scheme nor a converse was known in this case. It is this open problem that motivates this work.

The first contribution of this work is to show that the capacity of PIR-PSI is $C_{\text{\tiny PIR-PSI}}=\Psi(1/N, K-M)$, for arbitrary number of databases $N$, thus settling this open problem. As a generalization, we show that the capacity of TPIR-PSI, i.e., PIR-PSI where up to $T$ databases may collude, is $C_{\text{\tiny TPIR-PSI}}=\Psi(T/N, K-M)$. Evidently, the effect of private side information on capacity is the same as if the number of messages in TPIR was reduced from $K$ to $K-M$ \cite{Sun_Jafar_TPIR}. Remarkably, this is also the  capacity if the side-information is globally known to all databases as well. 

As the second contribution of this work, we  characterize the capacity of STPIR-PSI, i.e., PIR with private side information with symmetric privacy and robustness against any $T$-colluding servers. We show $C_{\text{\tiny TPIR-PSI}}=\Psi(T/N, \infty)$, provided that the databases have access to common randomness (not available to the user) in the amount that is at least $T/(N-T)$ bits per queried message bit.  Otherwise, the capacity of STPIR-PSI is zero. Note that this is identical to the capacity of STPIR with no side information  \cite{Wang_Skoglund}.

The remainder of this paper is organized as follows. Section \ref{sec:prob_stat} presents the problem statements. Section \ref{sec:result} presents the main results, i.e., the capacity characterizations  of TPIR-PSI and STPIR-PSI. The proofs of the capacity results are presented in Section \ref{sec:T1} and Section \ref{sec:T2}, and we  conclude with Section \ref{sec:conclusion}. 

{\it Notation:} We use bold font for random variables to distinguish them from deterministic variables, that are shown in normal font. For integers $z_1<z_2$, $[z_1:z_2]$ represents the set $\{z_1,z_1+1,\cdots, z_2\}$ and $(z_1:z_2)$ represents the vector $(z_1,z_1+1,\cdots, z_2)$. The compact notation $[z]$ represents $[1:z]$ for positive integer $z$. For random variables $\bm{W}_i, i=1,2,\dots,$ and a set of positive integers $S = \{s_1,s_2,\cdots,s_n\}$, where $s_1<s_2<\cdots<s_n$, the notation $\bm{{W}}_S$ represents the vector $(\bm{W}_{s_1},\bm{W}_{s_2},\cdots,\bm{W}_{s_n})$. For a matrix $G$ and a vector $S$, the notation $G[S,:]$ represents the submatrix of $G$ formed by retaining only the rows corresponding to the elements of the vector $S$. 
$o(L)$ represents a function of $L$ such that $\lim_{L\rightarrow\infty}o(L)/L=0$. $\mathbb{F}_q$ represents the finite field of size $q$.

\section{Problem Statements}\label{sec:prob_stat}
\subsection{TPIR-PSI: $T$-Private Information Retrieval with Private Side Information}
The TPIR-PSI problem is parametrized by $(K, M, N, T)$. Consider $K$ independent messages $\bm{{W}}_{[K]}=(\bm{W}_{1}, \cdots, \bm{W}_{K})$, each containing $L$ independent and uniform bits, i.e.,
\begin{eqnarray}
& H(\bm{W}_{1}, \cdots, \bm{W}_{K}) = H(\bm{W}_{1}) + \cdots + H(\bm{W}_{K}),\\
& H(\bm{W}_{1}) = \cdots = H(\bm{W}_{K}) = L. \label{size}
\end{eqnarray}
There are $N$ databases and each database stores all  $K$ messages $\bm{W}_{1}, \cdots, \bm{W}_{K}$. A user is equipped with a local cache and has $M$ $(M<K)$ messages as side information. Let $\bm{S}=\{\bm{i}_1, \bm{i}_2, \cdots, \bm{i}_M\}$ be $M$ distinct indices chosen uniformly from $[K]$. These $M$ cached messages are represented as $\bm{{W}}_{\bm{S}}=(\bm{W}_{\bm{i}_{1}}, \cdots, \bm{W}_{\bm{i}_{M}})$. $\bm{S}$ is not known to the databases. A user wishes to retrieve $\bm{W}_{\bm{\Theta}},$ where  $\bm{\Theta}$ is a message index uniformly chosen from $[K]\backslash \bm{S}$, as efficiently as possible, while revealing no information about $(\bm{\Theta}, \bm{S})$ to any colluding subsets of up to $T$ out of the $N$ databases. Note the following independence,
\begin{eqnarray}
H(\bm{\Theta}, \bm{S}, \bm{W}_1,\bm{W}_2, \cdots, \bm{W}_K)&=& H(\bm{\Theta},\bm{S})+H(\bm{W}_1)+\cdots+H(\bm{W}_K).
\end{eqnarray}

In order to retrieve $\bm{W}_{\bm{\Theta}}$, the user generates $N$ queries $\bm{Q}_{1}^{[\bm{\Theta},\bm{S}]}, \cdots, \bm{Q}_{N}^{[\bm{\Theta},\bm{S}]}$ with knowledge of $(\bm{\Theta},\bm{S}, \bm{W}_{\bm S})$. Since the queries are generated with no knowledge of the other $K-M$ messages, the queries must be independent of them,
\begin{eqnarray}
I\left(\bm{\Theta, S}, \bm{W}_{\bm{S}}, \bm{Q}_{1}^{[\bm{\Theta},\bm{S}]}, \cdots, \bm{Q}_{N}^{[\bm{\Theta},\bm{S}]} ; \bm{{W}}_{[K] \backslash \bm{S}} \right) = 0. \label{query}
\end{eqnarray}
The user sends query $\bm{Q}_{n}^{[\bm{\Theta},\bm{S}]}$ to the $n$-th database and in response, the $n$-th database returns an answer  $\bm{A}_{n}^{[\bm{\Theta},\bm{S}]}$ which is a deterministic function of $\bm{Q}_{n}^{[\bm{\Theta},\bm{S}]}$ and $\bm{{W}}_{[K]}$, 
\begin{align}
H\left(\bm{A}_n^{[\bm{\Theta},\bm{S}]}   \mid   \bm{Q}_n^{[\bm{\Theta},\bm{S}]}, \bm{W}_1, \cdots, \bm{W}_K  \right) = 0.\label{eq:ans}
\end{align}
Upon collecting the answers from all  $N$ databases, the user must be able to decode the desired message $\bm{W}_{\bm{\Theta}}$ based on the queries and side information, subject to a probability of error $P_e$ which must approach zero as the size of each message, $L$, approaches infinity. This is called the ``correctness'' constraint. From Fano's inequality, we have
\begin{eqnarray}
\mbox{[Correctness]} ~H\left(\bm{W}_{\bm{\Theta}}  \mid  \bm{A}_n^{[\bm{\Theta},\bm{S}]}, \bm{Q}_n^{[\bm{\Theta},\bm{S}]}, \bm{{W}}_{\bm{S}}, \bm{S}, \bm{\Theta}\right) = o(L). \label{correct:PSI}
\end{eqnarray} 

To satisfy the user-privacy constraint that any $T$ colluding databases learn nothing about $(\bm{\Theta}, \bm{S})$, the information available to any $T$ databases (queries, answers and stored messages) must be independent of $(\bm{\Theta}, \bm{S}).$ \footnote{Note that  the joint privacy of $(\bm{\Theta, S})$ is a stronger constraint than the marginal privacy of each of $\bm{\Theta}$ and $\bm{S}$, i.e., $I(\bm{\Theta, S}; \bm{Q}_{\mathcal{T}}^{[\bm{\Theta,S}]}, \bm{A}_{\mathcal{T}}^{[\bm{\Theta,S}]}, \bm{{W}}_{[K]})=0$ implies both $I(\bm\Theta; \bm{Q}_{\mathcal{T}}^{[\bm{\Theta,S}]}, \bm{A}_{\mathcal{T}}^{[\bm{\Theta,S}]}, \bm{{W}}_{[K]})=0$ and $I(\bm{S}; \bm{Q}_{\mathcal{T}}^{[\bm{\Theta,S}]}, \bm{A}_{\mathcal{T}}^{[\bm{\Theta,S}]}, \bm{{W}}_{[K]})=0$. However, the reverse is not true, i.e., even if both $I(\bm\Theta; \bm{Q}_{\mathcal{T}}^{[\bm{\Theta,S}]}, \bm{A}_{\mathcal{T}}^{[\bm{\Theta,S}]}, \bm{{W}}_{[K]})=0$ and $I(\bm{S}; \bm{Q}_{\mathcal{T}}^{[\bm{\Theta,S}]}, \bm{A}_{\mathcal{T}}^{[\bm{\Theta,S}]}, \bm{{W}}_{[K]})=0$, this does not imply that $I(\bm{\Theta, S}; \bm{Q}_{\mathcal{T}}^{[\bm{\Theta,S}]}, \bm{A}_{\mathcal{T}}^{[\bm{\Theta,S}]}, \bm{{W}}_{[K]})=0$.} Let $\mathcal{T}$ be any subset of $[1:N]$, of cardinality $|\mathcal{T}|=T$. $\bm{Q}_{\mathcal{T}}^{[\bm{\Theta},\bm{S}]}$ represents the vector of queries corresponding to $\bm{Q}_{n}^{[\bm{\Theta},\bm{S}]},  n \in \mathcal{T}$. $\bm{A}_{\mathcal{T}}^{[\bm{\Theta},\bm{S}]}$ is defined as the answer vector corresponding to $\bm{A}_{n}^{[\bm{\Theta},\bm{S}]} ,  n \in \mathcal{T}$. To satisfy the $T$-privacy requirement we must have 
\begin{align}
\mbox{[User privacy]} ~~~ I\left(\bm{\Theta},\bm{S}; \bm{Q}_{\mathcal{T}}^{[\bm{\Theta},\bm{S}]}, \bm{A}_{\mathcal{T}}^{[\bm{\Theta}, \bm{S}]}, \bm{{W}}_{[K]}\right)=0,&& \forall \mathcal{T} \subset [1:N], |\mathcal{T}|=T.  \label{privacy} 
\end{align} 

A TPIR-PSI scheme is called \emph{feasible} if it satisfies  the correctness constraint (\ref{correct:PSI}) and the user-privacy constraint (\ref{privacy}). For a feasible scheme, the TPIR-PSI rate characterizes asymptotically how many bits of desired information are retrieved per downloaded bit, and is defined as follows.
\begin{eqnarray}
R_{\text{\tiny TPIR-PSI}} \triangleq \lim_{L\rightarrow\infty}\frac{L}{D} ,
\end{eqnarray}
where $D$ is the expected (over all $\bm{\Theta}$ and $\bm{S}$) total number of bits downloaded by the user from all the databases. The \emph{capacity}, $C_{\text{\tiny TPIR-PSI}}$, is the supremum of $R_{\text{\tiny TPIR-PSI}}$ over all feasible schemes. 

\subsection{STPIR-PSI: Symmetric $T$-Private Information Retrieval with Private Side Information}
In symmetric $T$-colluding private information retrieval, an additional constraint is imposed: database privacy, which means that the user does not learn any information about $\bm{W}_{[K]}$ beyond the retrieved message, $\bm{W}_{\bm{\Theta}}$. To facilitate database privacy, suppose the databases share a common random variable $\bm{U}$ that is not known to the user. It has been shown that without such common randomness, symmetric PIR is not feasible when there is more than one message \cite{SymPIR, Sun_Jafar_SPIR}. The common randomness is independent of the messages, the desired messages index, and the side information index, so that 
\begin{eqnarray}
H\left(\bm{\Theta}, \bm{S}, \bm{W}_1, \cdots, \bm{W}_K, \bm{U}\right)  = H\left(\bm{\Theta}, \bm{S}\right) +  H\left(\bm{W}_1\right) + \cdots + H\left(\bm{W}_K \right) + H(\bm{U}). \label{indep}
\end{eqnarray}
The answering string $\bm{A}_{n}^{[\bm{\Theta},\bm{S}]}$ is a deterministic function of $\bm{Q}_{n}^{[\bm{\Theta},\bm{S}]}$, $\bm{{W}}_{[K]}$ and common randomness $\bm{U}$,
\begin{eqnarray}
H\left(\bm{A}_n^{[\bm{\Theta},\bm{S}]}  \mid  \bm{Q}_n^{[\bm{\Theta},\bm{S}]},  \bm{W}_1, \cdots, \bm{W}_K, \bm{U}\right) = 0. \label{ansdet}
\end{eqnarray}
The correctness condition is  the same as (\ref{correct:PSI}). The user-privacy condition is 
\begin{align}
\mbox{[User privacy]} ~~~ I\left(\bm{\Theta},\bm{S}; \bm{Q}_{\mathcal{T}}^{[\bm{\Theta},\bm{S}]}, \bm{A}_{\mathcal{T}}^{[\bm{\Theta}, \bm{S}]}, \bm{\bm{W}}_{[K]}, \bm{U}\right)=0, &&\forall \mathcal{T} \subset [1:N], |\mathcal{T}|=T.  \label{privacy:symmetric} 
\end{align} 
Database privacy requires that the user learns nothing  about $\bm{{W}}_{\overline{(\bm{\Theta,S})}} = \bm{{W}}_{[K]\backslash(\{\bm{\Theta}\}\cup\bm{S})}$, i.e., messages other than his desired message and side-information. Therefore, 
\begin{eqnarray}
\mbox{[Database privacy]} ~~ I\left(\bm{{W}}_{\overline{(\bm{\Theta,S})}}~;~ \bm{Q}_{[N]}^{[\bm{\Theta},\bm{S}]}, \bm{A}_{[N]}^{[\bm{\Theta},\bm{S}]} ,  \bm{\Theta}, \bm{S}, \bm{{W}}_{\bm{S}}\right) = 0. \label{noleak}
\end{eqnarray}

An STPIR-PSI scheme is called \emph{feasible} if it satisifes the correctness constraint (\ref{correct:PSI}), the user-privacy constraint (\ref{privacy:symmetric}) and the database-privacy constraint (\ref{noleak}). For a feasible scheme, the STPIR-PSI rate characterizes how many bits of desired information are retrieved per downloaded bit. The \emph{capacity}, $C_{\text{\tiny STPIR-PSI}}$, is the supremum of  rates over all feasible STPIR-PSI schemes.  

\section{Main Results} \label{sec:result}
The following theorem presents our first result, the capacity of TPIR-PSI.

\begin{theorem}\label{thm:1}
For the TPIR-PSI problem with $K$ messages, $N$ databases and $M$ $(M<K)$ side information messages, the capacity is
\begin{eqnarray}
C_{\text{\tiny TPIR-PSI}} = \left(1+\frac{T}{N}+\left(\frac{T}{N}\right)^2+\cdots+\left(\frac{T}{N}\right)^{K-M-1}\right)^{-1}.\label{eq:ctpirpsi}
\end{eqnarray}
\end{theorem}

The following observations place Theorem \ref{thm:1} in perspective.

\begin{enumerate}
\item The expression $C_{\text{\tiny TPIR-PSI}}$ equals  the capacity of TPIR with $K-M$ messages \cite{Sun_Jafar_TPIR}. Evidently, the impact of private side information is equivalent to reducing the effective number of messages from $K$ to $K-M$. 

\item The capacity result stays the same if the side information is globally known, i.e., not only the user, but also the databases are knowledgeable about the side information.\footnote{If the $M$ side information messages are globally known, the capacity is $C_g = \left(1+\frac{T}{N}+\cdots+\left(\frac{T}{N}\right)^{K-M-1}\right)^{-1}$. The achievable scheme is the TPIR scheme of \cite{Sun_Jafar_TPIR} after the cached messages are eliminated. The converse bound can be proved as follows. Suppose the capacity is greater than $C_g$, then there is a scheme $\Pi$ that achieves a  larger rate than $C_g$ in the presence of the $M$ globally known messages. Consider a TPIR problem with $K-M$ messages. It can be  assumed that there are $M$ globally known dummy messages. With this globally known side information, the user can use the scheme $\Pi$ to retrieve the message and to achieve a rate larger than $C_g$, which violates the capacity of TPIR. Therefore, the capacity of TPIR with globally known side information is $C_g$.} It is remarkable that the capacity of PIR with public side information is the same as the capacity of PIR with private side information, but not the same as the capacity of the intermediate setting  where each of $\bm{\Theta}$ and $\bm{S}$ is private, but  $(\bm{\Theta,S})$ are not \emph{jointly} private. It is easy to find examples (see \cite{Kadhe_Garcia_Heidarzadeh_Rouayheb_Sprintson}) even with only one database where the achievable rate for the latter setting exceeds the capacity in \eqref{eq:ctpirpsi}. For example consider $N=1$ database that stores $K=4$ messages, of which $M=1$ message is available to the user as side information. For PIR-PSI, where the joint privacy of $(\bm{\Theta,S})$ must be preserved, the capacity is $1/3$. The capacity is also $1/3$ if ${\bf S}$ is public, i.e., known to the database. However, the PIR-NSI scheme in \cite{Kadhe_Garcia_Heidarzadeh_Rouayheb_Sprintson}, which preserves the privacy of $\bm{\Theta}$ and the privacy of $\bm{S}$ individually (but not their joint privacy), has capacity $1/2$ which is strictly larger.
\end{enumerate}
Our second result is the capacity of STPIR-PSI, presented in the following theorem.
\begin{theorem}\label{thm:2}
For the STPIR-PSI problem with $K\ge2$ messages, $N$ databases and $M$ $(M<K)$ side information messages, the capacity is
\begin{equation}
C_{\text{\tiny STPIR-PSI}}  =\left\{
\begin{array}{rcl}
1,& & \mbox{if} ~~M=K-1,\\
1 - \frac{T}{N}, & & \mbox{if} ~~M<K-1~~\mbox{and} ~~ \rho \geq \frac{T}{N-T},\\
0, & & \mbox{otherwise},\\
\end{array} \right.
\end{equation}

\noindent where $\rho  =\frac{H(\bm{U})}{L}$ is the amount of common randomness available to the databases, normalized by the message size.
\end{theorem}

The following observations are in order.
\begin{enumerate}
\item When there is only $K=1$ message, or when there are $M=K-1$ side information messages, the database-privacy constraint is satisfied trivially, so STPIR reduces to the TPIR setting and the capacity is $1$. Note that for symmetric PIR without side information, when $K \ge 2$, the common randomness is necessary for feasibility. However, for STPIR-PSI, if there are $M=K-1$ side information messages, then common randomness is not needed.

\item When $K\ge2$ and $M<K-1$, then $C_{\text{\tiny STPIR-PSI}}$ only depends on the number of databases $N$, the colluding parameter $T$, and the amount of common randomness. It is independent of the number of messages $K$ and the number of side information messages $M$. 

\item $C_{\text{\tiny STPIR-PSI}}$ is equal to  the capacity of STPIR without side information, which is the result of Theorem $1$ in \cite{Wang_Skoglund}. Namely, utilizing the side information does not help improve the capacity. In fact, in the achievable scheme, side information is not used to generate queries.

\item The capacity of STPIR-PSI is strictly smaller than the capacity of TPIR-PSI, which means that the additional requirement of preserving database privacy strictly penalizes the capacity. However, the penalty vanishes in the regime of large number of messages, i.e., $C_{\text{\tiny TPIR-PSI}} > C_{\text{\tiny STPIR-PSI}}$ for any finite $K$ and $C_{\text{\tiny TPIR-PSI}} \rightarrow C_{\text{\tiny STPIR-PSI}}$ when $K \rightarrow \infty$.
 
\item Theorem \ref{thm:2} can be also applied to the STPIR with globally known side information.\footnote{The explanation is similar to that for TPIR with globally known side information as in the previous footnote.}
\end{enumerate}

\section{Proof of Theorem \ref{thm:1}} \label{sec:T1}
\subsection{Achievability}
The backbone of the achievable scheme for TPIR-PSI with parameters $(K, M, N, T)$ is the achievable scheme of TPIR \cite{Sun_Jafar_TPIR}. We inherit the steps of the query structure construction and query structure specialization. The novel element of the achievable scheme is query redundancy removal based on the side information. To illustrate how this idea works, we present two examples with small values of $K, M, N$ and $T$, and then generalize it to arbitrary $K, M, N$ and $T$.

\subsubsection{Example with $(K, M, N, T)=(3, 1, 2, 1)$}
We start with the achievable scheme of PIR  ($T=1$). Suppose there are $N=2$ databases and $K=3$ messages, one of which is known to the user as side information. Let each message consist of $N^K=8$ symbols from a finite field $\mathbb{F}_q$ that is large enough so that a systematic maximum distance separable (MDS) code with length $13$ and dimension $7$ exists. A systematic code is a code in which the input data is embedded in the encoded output. According to the query structure construction and specialization of \cite{Sun_Jafar_TPIR}, the messages $\bm{W}_1, \bm{W}_2, \bm{W}_3 \in \mathbb{F}_q^{8}$ are represented as $8\times1$ column vectors over $\mathbb{F}_q$. Let $\bm{U}_1, \bm{U}_2, \bm{U}_3 \in \mathbb{F}_q^{8\times8}$ represent random matrices chosen privately by the user, independently and uniformly from all $8\times 8$ full-rank matrices over $\mathbb{F}_q$. Without loss of generality, suppose $(\bm{\Theta},\bm{S})=(1,\{3\})$, i.e., the user knows $\bm{W}_3$ as side information, and wishes to retrieve $\bm{W}_1$. Define the $8\times1$ column vectors $\bm{a}_{(1:8)}, \bm{b}_{(1:8)}, \bm{c}_{(1:8)} \in \mathbb{F}_q^{8}$ as below
\begin{align}
\bm{a}_{(1:8)} = \bm{U}_1\bm{W}_1,&&\bm{b}_{(1:8)} = \bm{U}_2\bm{W}_2, &&\bm{c}_{(1:8)} = \bm{U}_3\bm{W}_3. 
\end{align}
The downloaded symbols from each database are represented below. We use DB$_i$ to represent the $i^{th}$ database.
\begin{eqnarray*}
\begin{array}{ccc}
\begin{array}{|c|c|c|}\hline
\mbox{\small DB$_1$} & \mbox{\small DB$_2$} \\ \hline
\bm{a}_1, \bm{b}_1, \bm{c}_1 & \bm{a}_2, \bm{b}_2, \bm{c}_2 \\
\bm{a}_3 + \bm{b}_2& \bm{a}_5 + \bm{b}_1\\
\bm{a}_4 + \bm{c}_2& \bm{a}_6 + \bm{c}_1\\
\bm{b}_3 + \bm{c}_3& \bm{b}_4 + \bm{c}_4 \\
\bm{a}_7 + \bm{b}_4 +\bm{c}_4& \bm{a}_8 + \bm{b}_3 + \bm{c}_3\\
\hline
\end{array} 
\end{array}
\end{eqnarray*}
Note that the rate achieved by this scheme is $R_{\text{\tiny PIR}} = 4/7$ (which is also the capacity in the absence of side information), since the user recovers $8$ desired symbols from a total of $14$ downloaded symbols. 

However, in the PIR-PSI setting $\bm{W}_3$ is the side information, which means the user already knows $\bm{c}_{(1:8)}$. Specificallly consider the download from DB$_1$, where $\bm{c}_1$ is redundant. What the user needs now is only the remaining $6$ symbols from DB$_1$. So we take the step of query redundancy removal. The idea is that the user asks each database to encode the $7$ original symbols with a systematic $(13, 7)$ MDS code and downloads only the $6$ linear combinations corresponding to the non-systematic part, called parity symbols. Since the user already knows $\bm{c}_1$, along with the $6$ downloaded symbols, this gives him $7$ symbols of the MDS code, from which he can recover all $7$ symbols of the original PIR scheme. Formally, let $G^s_{e \times f}$ denote the generator matrix of a systematic $(e, f)$ MDS code (e.g., a Reed Solomon code). The generator matrix does not need to be random, i.e., it may be globally known. Suppose $G^s_{13 \times 7} = \left[V_{7\times 6} \mid I_{7\times 7}\right]'$, where $I_{7\times 7}$ is the identity matrix. Denote by vector $\bm{X}_i \in \mathbb{F}_q^{7}$ the symbols downloaded from DB$_i$ after the query structure construction and specialization. Therefore $\bm{X}_1=(\bm{a}_1,\bm{b}_1, \bm{c}_1, \bm{a}_3+\bm{b}_2, \bm{a}_4+\bm{c}_2, \bm{b}_3+\bm{c}_3, \bm{a}_7+\bm{b}_4+\bm{c}_4)$ and $\bm{X}_2=(\bm{a}_2, \bm{b}_2, \bm{c}_2, \bm{a}_5+\bm{b}_1, \bm{a}_6+\bm{c}_1, \bm{b}_4+\bm{c}_4, \bm{a}_8+\bm{b}_3+\bm{c}_3)$. With query redundancy removal, from DB$_1$ and DB$_2$ the user downloads $V'_{7\times 6}\bm{X}_1$ and $V'_{7\times 6}\bm{X}_2$ instead of $\bm{X}_1$, $\bm{X}_2$. The correctness constraint is satisfied because of the property of MDS code, i.e., given $\bm{c}_1, \bm{c}_2, V'_{7\times 6}\bm{X}_1$ and $V'_{7\times 6}\bm{X}_2$, the user is able to decode $\bm{X}_1$ and $\bm{X}_2$. The privacy is essentially inherited from the original PIR scheme and the fact that the MDS code is fixed \emph{a priori}, i.e., it does not depend on $(\bm\Theta,\bm{S})$. Thus, the rate achieved with private side information is $R_{\text{\tiny PIR-PSI}} =8/12=2/3=(1+1/2)^{-1}$ which matches the capacity for this setting.

\subsubsection{Example with $(K, M, N, T)=(3, 2, 3, 2)$}
The second example allows any $T=2$ databases to collude. Consider the case where $(K, M, N, T)=(3,2,3, 2)$, i.e., there are $3$ messages and $3$ databases. Assume the user knows $\bm{W}_2$ and $\bm{W}_3$ as side information and wishes to retrieve $\bm{W}_1$. Based on the query structure construction and specialization, let each message consist of $N^K=27$ symbols from a large enough finite field $\mathbb{F}_q$. The messages $\bm{W}_1, \bm{W}_2, \bm{W}_3 \in \mathbb{F}_q^{27}$ are $27 \times 1$ column vectors and $\bm{U}_1, \bm{U}_2, \bm{U}_3 \in \mathbb{F}_q^{27\times27}$ represent random matrices chosen privately by the user, independently and uniformly from all $27\times 27$ full-rank matrices over $\mathbb{F}_q$.  Let ${G}_{e\times f}$ denote the generator matrix of an $(e,f)$ MDS code (e.g., a Reed Solomon code). The generator matrices ${G}_{e\times f}$ need not be systematic or random, and may be globally known. Define the $27\times1$ column vectors $\bm{a}_{(1:27)}, \bm{b}_{(1:27)}, \bm{c}_{(1:27)} \in \mathbb{F}_q^{27}$ as follows.
\begin{eqnarray}
\bm{a}_{(1:27)} &=& \bm{U}_1\bm{W}_1,\\
\bm{b}_{(1:18)} &=& {G}_{18\times12}\bm{U}_2[(1:12),:]\bm{W}_2,\label{m1}\\
\bm{c}_{(1:18)} &=& {G}_{18\times12}\bm{U}_3[(1:12),:]\bm{W}_3,\label{m2}\\
\bm{b}_{(19:27)} &=& {G}_{9\times6}\bm{U}_2[(13:18),:]\bm{W}_2,\label{m3}\\
\bm{c}_{(19:27)} &=& {G}_{9\times6}\bm{U}_3[(13:18),:]\bm{W}_3,\label{m4}
\end{eqnarray} 
where $\bm{U}_2[(1:18),:]$ and $\bm{U}_3[(1:18),:]$ are $18\times27$ matrices comprised of the first $18$ rows of $\bm{U}_2$ and $\bm{U}_3$, respectively. Note that the same generator matrix ${G}_{18\times12}$ is used in (\ref{m1}) and (\ref{m2}), and the same generator matrix ${G}_{9\times6}$ is used in (\ref{m3}) and (\ref{m4}). The scheme below correctly recovers the queried message and maintains user privacy even if $2$ databases collude \cite{Sun_Jafar_TPIR}.
\begin{eqnarray*}
\begin{array}{cccc}
\begin{array}{|c|c|c|c|}\hline
\mbox{\small DB$_1$} & \mbox{\small DB$_2$} & \mbox{\small DB$_3$}  \\ \hline
\bm{a}_1, \bm{a}_2, \bm{a}_3, \bm{a}_4& \bm{a}_5, \bm{a}_6, \bm{a}_7, \bm{a}_8 & \bm{a}_9, \bm{a}_{10}, \bm{a}_{11}, \bm{a}_{12} \\
\bm{b}_1, \bm{b}_2, \bm{b}_3, \bm{b}_4& \bm{b}_5, \bm{b}_6, \bm{b}_7, \bm{b}_8 & \bm{b}_9, \bm{b}_{10}, \bm{b}_{11}, \bm{b}_{12} \\
\bm{c}_1, \bm{c}_2, \bm{c}_3, \bm{c}_4& \bm{c}_5, \bm{c}_6, \bm{c}_7, \bm{c}_8 & \bm{c}_9, \bm{c}_{10}, \bm{c}_{11}, \bm{c}_{12} \\
\bm{a}_{13} + \bm{b}_{13}& \bm{a}_{15} + \bm{b}_{15} & \bm{a}_{21}+\bm{b}_{17} \\
\bm{a}_{14} + \bm{b}_{14}& \bm{a}_{16} + \bm{b}_{16} & \bm{a}_{22}+\bm{b}_{18} \\
\bm{a}_{17} + \bm{c}_{13}& \bm{a}_{19} + \bm{c}_{15} & \bm{a}_{23}+\bm{c}_{17} \\
\bm{a}_{18} + \bm{c}_{14}& \bm{a}_{20} + \bm{c}_{16} & \bm{a}_{24}+\bm{c}_{18} \\
\bm{b}_{19} + \bm{c}_{19}& \bm{b}_{21} + \bm{c}_{21} & \bm{b}_{23}+\bm{c}_{23} \\
\bm{b}_{20} + \bm{c}_{20}& \bm{b}_{22} + \bm{c}_{22} & \bm{b}_{24}+\bm{c}_{24} \\
\bm{a}_{25} + \bm{b}_{25} + \bm{c}_{25}& \bm{a}_{26} + \bm{b}_{26} + \bm{c}_{26} & \bm{a}_{27} + \bm{b}_{27} + \bm{c}_{27}\\
\hline
\end{array} 
\end{array}
\end{eqnarray*}
Without side information, the achieved rate is $R_{\text{\tiny TPIR}} = 9/19$. Note that in this scheme the user downloads $19$ symbols from each database. Now we take the step of query redundancy removal. Since the user knows $\bm{W}_2, \bm{W}_3$ as side information, he does not need to download symbols of $\bm{W}_2, \bm{W}_3$, or the linear combinations comprised of only  $\bm{W}_2, \bm{W}_3$ symbols, i.e., $\bm{b}_i, \bm{c}_i, 1 \le i \le 12$ and $\bm{b}_j+\bm{c}_j, 19 \le j \le 24$ from the databases. Therefore $10$ redundant symbols may be reduced from each database. Denote by vector $\bm{X}_i \in \mathbb{F}_q^{19}$ the symbols downloaded from DB$_i$ after the query structure construction and specialization. The user asks each database to encode $\bm{X}_i$ with a systematic $(28, 19)$ MDS code generator matrix $G^s_{28 \times 19} = \left[V_{19\times 9} \mid I_{19\times 19}\right]'$ and downloads only the $9$ linear combinations corresponding to the parity part, $V'_{19\times 9}\bm{X}_i$. Consider DB$_1$. Since the user already knows $\bm{b}_i, \bm{c}_i, 1 \le i \le 4$ and $\bm{b}_j+\bm{c}_j, 19 \le j \le 20$, along with the $9$ downloaded symbols, this gives him $19$ symbols of the MDS code, from which he can recover all $19$ symbols of the original TPIR scheme. The correctness and privacy are inherited from the original scheme and the property of MDS code. The achieved rate is $R_{\text{\tiny TPIR-PSI}} =27/27=1$, which matches the capacity for this setting. 

\subsubsection{Arbitrary $(K, M, N, T)$}
{\bf Scheme description.} Now we are ready to present the achievable scheme for arbitrary $(K, M, N, T)$. For each database, let $p_1$ denote the number of symbols to be downloaded in the original TPIR scheme of \cite{Sun_Jafar_TPIR}.  Out of these $p_1$ symbols, let $p_2$ $(p_2 < p_1)$ denote the number of symbols that are already known to the user based on his side information.  Denote by vector $\bm{X}_i \in \mathbb{F}_q^{p_1}$ the symbols downloaded from DB$_i$ after the query structure construction and specialization according to \cite{Sun_Jafar_TPIR}. For each database, use a systematic $(2p_1-p_2, p_1)$ MDS code with generator matrix $G^s_{(2p_1-p_2) \times p_1} = \left[V_{p_1\times (p_1-p_2)} \mid I_{p_1\times p_1}\right]'$ to encode the $p_1$ symbols into $2p_1-p_2$ symbols, of which $p_1$ are systematic, and download only the $p_1-p_2$ parity symbols, $V'_{p_1\times (p_1-p_2)}\bm{X}_i$. 

Note that the user does not need to know the realization of side information $\bm{S}$ or $\bm{W}_{\bm{S}}$ in order to construct the queries. This is because the systematic MDS code in the query redundancy removal does not depend on $\bm{S}$ or $\bm{W}_{\bm S}$. During the decoding, $\bm{S}$ and $\bm{W}_{\bm S}$ are only used after the answers from the databases are collected. Therefore, the privacy of this TPIR-PSI scheme is inherited from the privacy of the original TPIR scheme. Correctness follows from the MDS property because in addition to the $p_1-p_2$ downloaded symbols from DB$_i$, i.e., $V'_{p_1\times (2p_1-p_2)}\bm{X}_i$, the user provides the $p_2$ symbols that he already knows, to obtain a total of $p_1$ symbols from the MDS code. Since any $p_1$ symbols from an MDS code suffice to recover the original $p_1$ symbols, the user recovers $\bm{X}_i$. Then the correctness is inherited from the correctness of the original TPIR scheme. All that remains is to calculate the rate achieved by this scheme.

{\bf Rate calculation.} For the TPIR scheme of \cite{Sun_Jafar_TPIR}, after the query structure construction and specialization, the response from each database is comprised of $K$ layers. Over the $k$-th layer, the downloads are in the form of sums of $k$ symbols, each from one distinct message, called $k$-sum. Note that there are $\binom{K}{k}$ possible ``types" of $k$-sums and $(N-T)^{k-1}T^{K-k}$ distinct instances\footnote{The term $(N-T)^{k-1}T^{K-k}$ comes from the undesired message exploitation step (Step 4) of achievability in \cite{Sun_Jafar_TPIR} and can be verified recursively. A detailed analysis in similar flavor can be found in \cite{Sun_Jafar_PIR}.} of each type of $k$-sum in $k$-th layer. So the total number of elements contained in layer $k$ is $\binom{K}{k}(N-T)^{k-1}T^{K-k}$. Therefore, the total number of symbols to be downloaded from each database is  $p_1=\sum_{k=1}^K\binom{K}{k}(N-T)^{k-1}T^{K-k}$. 

The next step is to calculate, out of these $p_1$ symbols, how many are already known to the user based on his side information. Suppose the user knows the $M$ messages $\bm{W}_{\bm{i}_1}, \cdots, \bm{W}_{\bm{i}_M}$, $\{\bm{i}_{1}, \cdots, \bm{i}_{M}\}\in[K]$ as side information beforehand. Thus the user knows all linear combinations that are comprised of symbols from these $M$ messages. In terms of layer $k$ $(k \leq M)$, the user knows all the instances of $k$-sum that contain only symbols   $\bm{W}_{\bm{j}_1},\bm{W}_{\bm{j}_2},\cdots,\bm{W}_{\bm{j}_k}$, where $\{\bm{j}_1,\bm{j}_2,\cdots,\bm{j}_k\}\subset \{\bm{i}_1, \cdots,\bm{i}_M\}$. So the total number of symbols known to the user corresponding to each database is $p_2=\sum_{k=1}^M\binom{M}{k}(N-T)^{k-1}T^{K-k}$. 
\allowdisplaybreaks
Notice that $p_1$ can be simplified as,
\begin{eqnarray}
p_1 &=& \sum_{k=1}^K (N-T)^{k-1} T^{K-k} \binom{K}{k} \\
&=& \frac{ \sum_{k=0}^{K} (N-T)^k T^{K-k} \binom{K}{k} - T^K }{N-T} \\
&=& \frac{ N^K - T^K }{N-T}.
\end{eqnarray}
And $p_2$ can be simplified as,
\begin{eqnarray}
p_2 &=& \sum_{k=1}^M (N-T)^{k-1} T^{K-k} \binom{M}{k} \\
&=& T^{K-M} \sum_{k=1}^M (N-T)^{k-1} T^{M-k} \binom{M}{k} \\
&=& \frac{T^{K-M}(N^M-T^M)}{N-T}.
\end{eqnarray}
From each database the number of downloaded symbols of desired messages can be calculated as, 
\begin{eqnarray}
m = \sum_{k=1}^K (N-T)^{k-1} T^{K-k} \binom{K-1}{k-1} = N^{K-1}.
\end{eqnarray}
Therefore, the rate achieved is 
\begin{align}
R_{\text{\tiny TPIR-PSI}} &= \frac{Nm}{N(p_1-p_2)} \\
&= \frac{N^{K-1} (N-T)}{(N^K - T^K)- T^{K-M}(N^M-T^M)} \\
&= \frac{1-\frac{T}{N}}{1-(\frac{T}{N})^{K-M}}\\
& = \left( 1+\frac{T}{N} + \cdots + \left(\frac{T}{N}\right)^{K-M-1} \right)^{-1}.\\
\end{align}
This matches the capacity of TPIR-PSI, thus completing the proof of achievability for Theorem \ref{thm:1}.

\subsection{Converse}
Before presenting the general converse, let us start with a simple example for ease of exposition.
\subsubsection{Converse for $(K, M, N, T)=(3, 1, 2, 1)$}
Let $\mathcal{S}$ be a set whose elements are all possible realizations of $\bm{S}$, i.e., $\mathcal{S}=\{S ~\big| ~S\subset[K], |S|=M\}$. 
The total download is bounded as,
\begin{align}
D&\geq H({\bm A}_{[N]}^{[\bm{\Theta},{\bm S}]}  )\\
&\geq H({\bm A}_{[N]}^{[\bm{\Theta},{\bf S}]}\mid {\bm Q}_{[N]}^{[\bm{\Theta},\bm{S}]},\bm{{W}}_{\bm{S}},\bm{\Theta},\bm{S})\\
&\geq \min_{S\in\mathcal{S}, \theta\in[K]\backslash S}H({\bm A}_{[N]}^{[\bm{\Theta},{\bf S}]}\mid {\bm Q}_{[N]}^{[\bm{\Theta},\bm{S}]},\bm{{W}}_{\bm{S}},\bm{\Theta}=\theta,\bm{S}=S).\label{eq:minH}
\end{align}
\noindent We will derive a lower bound on the entropy in \eqref{eq:minH} that holds  for all $(\theta,S)$. 
For $(K, M, N, T)=(3, 1, 2, 1)$, without loss of generality suppose message $\bm{W}_1$  is known as side-information and $\bm{W}_2$ is desired. We bound the total download as,
\begin{align}
D&\geq H\left(\bm{A}_1^{[\bm{\Theta},\bm{S}]}, \bm{A}_2^{[\bm{\Theta},\bm{S}]} \mid \bm{Q}_1^{[\bm{\Theta},\bm{S}]}, \bm{Q}^{[\bm{\Theta},\bm{S}]}_2, \bm{W}_1,\bm{\Theta}=2,\bm{S}=\{1\}\right)\\
&\overset{(\ref{correct:PSI})}{=}H\left(\bm{A}_1^{[\bm{\Theta},\bm{S}]}, \bm{A}_2^{[\bm{\Theta},\bm{S}]}, \bm{W}_2 \mid \bm{Q}_1^{[\bm{\Theta},\bm{S}]}, \bm{Q}^{[\bm{\Theta},\bm{S}]}_2, \bm{W}_1,\bm{\Theta}=2,\bm{S}=\{1\}\right)+o(L)\\
&=H\left(\bm{W}_2 \mid \bm{Q}_1^{[\bm{\Theta},\bm{S}]}, \bm{Q}^{[\bm{\Theta},\bm{S}]}_2,  \bm{W}_1,\bm{\Theta}=2,\bm{S}=\{1\}\right) \nonumber \\
&\hspace{1cm}+H\left(\bm{A}_1^{[\bm{\Theta},\bm{S}]}, \bm{A}_2^{[\bm{\Theta},\bm{S}]} \mid  \bm{Q}_1^{[\bm{\Theta},\bm{S}]}, \bm{Q}^{[\bm{\Theta},\bm{S}]}_2, \bm{W}_1, \bm{W}_2,\bm{\Theta}=2,\bm{S}=\{1\}\right)+o(L)\\
&\geq L+H\left(\bm{A}_1^{[\bm{\Theta},\bm{S}]} \mid  \bm{Q}_1^{[\bm{\Theta},\bm{S}]}, \bm{Q}^{[\bm{\Theta},\bm{S}]}_2,  \bm{W}_1, \bm{W}_2,\bm{\Theta}=2,\bm{S}=\{1\}\right)+o(L)\label{eq:a}\\
&=L+H\left(\bm{A}_1^{[\bm{\Theta},\bm{S}]} \mid  \bm{Q}_1^{[\bm{\Theta},\bm{S}]}, \bm{W}_1, \bm{W}_2,\bm{\Theta}=2,\bm{S}=\{1\}\right)+o(L)\label{eq:mc}\\
&\overset{(\ref{privacy})}{=}L+H\left(\bm{A}_1^{[\bm{\Theta},\bm{S}]} \mid   \bm{Q}_1^{[\bm{\Theta},\bm{S}]},\bm{W}_1,\bm{W}_2,\bm{\Theta}=3,\bm{S}=\{1\}\right)+ o(L)
\label{eq:one}
\end{align}
where (\ref{eq:a}) holds because of (\ref{size}), (\ref{query}), the chain rule and non-negativity of entropy. Equation (\ref{eq:mc}) holds since given $\bm{\Theta}=2,\bm{S}=\{1\}$, we know that  $\bm{A}_1^{[\bm{\Theta},\bm{S}]}\leftrightarrow \left(\bm{Q}_1^{[\bm{\Theta},\bm{S}]}, \bm{W}_1, \bm{W}_2\right)\leftrightarrow \bm{Q}_2^{[\bm{\Theta},\bm{S}]}$ is a Markov chain.\footnote{This is easily verified because by the chain rule of mutual information, $I(\bm{A}_1^{[\bm{\Theta},\bm{S}]}; \bm{Q}_2^{[\bm{\Theta},\bm{S}]}\mid \bm{Q}_1^{[\bm{\Theta},\bm{S}]}, \bm{W}_1, \bm{W}_2,\bm{\Theta}=2,\bm{S}=\{1\})=I(\bm{A}_1^{[\bm{\Theta},\bm{S}]}; \bm{Q}_2^{[\bm{\Theta},\bm{S}]}\mid \bm{Q}_1^{[\bm{\Theta},\bm{S}]}, \bm{W}_{\{1,2,3\}},\bm{\Theta}=2,\bm{S}=\{1\})+I(\bm{W}_{3}; \bm{Q}_2^{[\bm{\Theta},\bm{S}]}\mid \bm{Q}_1^{[\bm{\Theta},\bm{S}]}, \bm{W}_1, \bm{W}_2,\bm{\Theta}=2,\bm{S}=\{1\})-I(\bm{W}_3; \bm{Q}_2^{[\bm{\Theta},\bm{S}]}\mid \bm{Q}_1^{[\bm{\Theta},\bm{S}]}, \bm{W}_1, \bm{W}_2,\bm{A}_1^{[\bm{\Theta},\bm{S}]}, \bm{\Theta}=2,\bm{S}=\{1\}).$ The first two mutual information terms on the right hand side are equal to zero because of \eqref{eq:ans} and \eqref{query}, respectively, leaving only the negative mutual information term which must also be zero because mutual information cannot be negative.}

Similarly,
\begin{eqnarray}
D&\geq&  L+H\left(\bm{A}_2^{[\bm{\Theta},\bm{S}]} \mid  \bm{Q}_2^{[\bm{\Theta},\bm{S}]}, \bm{W}_1, \bm{W}_2,\bm{\Theta}=3,\bm{S}=\{1\}\right)+o(L).
\label{eq:two}
\end{eqnarray}

Adding (\ref{eq:one}) and (\ref{eq:two}) we have
\begin{align}
2D &\geq 2L  +H\left(\bm{A}_1^{[\bm{\Theta},\bm{S}]} \mid  \bm{W}_1,\bm{W}_2,\bm{Q}_1^{[\bm{\Theta},\bm{S}]},\bm{\Theta}=3,\bm{S}=\{1\}\right) \nonumber \\
&\hspace{1cm}+H\left(\bm{A}_2^{[\bm{\Theta},\bm{S}]} \mid  \bm{W}_1,\bm{W}_2,\bm{Q}_2^{[\bm{\Theta},\bm{S}]},\bm{\Theta}=3,\bm{S}=\{1\}\right)+ o(L)\\
&\geq 2L +H\left(\bm{A}_1^{[\bm{\Theta},\bm{S}]},\bm{A}_2^{[\bm{\Theta},\bm{S}]} \mid \bm{W}_1, \bm{W}_2, \bm{Q}_1^{[\bm{\Theta},\bm{S}]}, \bm{Q}_2^{[\bm{\Theta},\bm{S}]},\bm{\Theta}=3,\bm{S}=\{1\}\right)+ o(L)\\
&\geq 2L +H\left(\bm{W}_3 \mid \bm{W}_1, \bm{W}_2, \bm{Q}_1^{[\bm{\Theta},\bm{S}]}, \bm{Q}_2^{[\bm{\Theta},\bm{S}]},\bm{\Theta}=3,\bm{S}=\{1\}\right)+o(L)\label{eq:b}\\
&= 2L +L +o(L)\\
&= 3L+ o(L),
\end{align}
Here (\ref{eq:b}) holds because from $\left(\bm{W}_1, \bm{W}_2, \bm{Q}_1^{[\bm{\Theta},\bm{S}]}, \bm{Q}_2^{[\bm{\Theta},\bm{S}]},\bm{\Theta}=3,\bm{S}=\{1\}\right)$ one can recover $\bm{W}_3$ with vanishing probability of error. Since the same argument holds for all realizations $(\bm{\Theta,S})=(\theta,S)$, 
this gives us the bound on the capacity of TPIR-PSI with $(K,M,N,T)=(3,1,2,1)$ as $C_{\text{\tiny TPIR-PSI}} \leq \frac{2}{3}$.

\subsubsection{Converse for Arbitrary $(K, M, N, T)$}
If $M=K-1$, then the capacity is $1$ which is trivial. So let us assume that $M<K-1$. For compact notation, let us define 
\begin{eqnarray}
D(K, S,\theta, V) &\triangleq&  H\left(\bm{A}_{[N]}^{[\bm{\Theta},\bm{S}]}\mid{\bm Q}_{[N]}^{[\bm{\Theta},\bm{S}]}, \bm{W}_{[V]},\bm{\Theta}=\theta, \bm{S}=S\right).
\end{eqnarray}
Here   $\bm{W}_{[V]}=(\bm{W}_1,\bm{W}_2,\cdots,\bm{W}_V)$ represents the messages that appear in the conditioning. Also, define an arbitrary $\mathcal{T} \subset [N]$ with cardinality $|\mathcal{T}|=T$ which represents the set of  indices of colluding databases. 

Without loss of generality, let us consider $\bm{S}=[M]$, $\bm{\Theta}=M+1$. Then, we have
\begin{align}
&D(K,[M],M+1,M)\\
&=H(\bm{A}_{[N]}^{[\bm{\Theta},\bm{S}]}|\bm{Q}_{[N]}^{[\bm{\Theta},\bm{S}]}, \bm{W}_{[M]},\bm{\Theta}=M+1,\bm{S}=[M])\\
&\overset{(\ref{correct:PSI})}{=} H\left(\bm{A}_{[N]}^{[\bm{\Theta},\bm{S}]}, \bm{W}_{M+1}\mid \bm{Q}_{[N]}^{[\bm{\Theta},\bm{S}]}, \bm{W}_{[M]},\bm{\Theta}=M+1,\bm{S}=[M]\right)+ o(L)\\
&= H\left(\bm{W}_{M+1}\mid\bm{Q}_{[N]}^{[\bm{\Theta},\bm{S}]}, \bm{W}_{[M]},\bm{\Theta}=M+1,\bm{S}=[M]\right) \nonumber\\
&\hspace{1cm}+H\left(\bm{A}_{[N]}^{[\bm{\Theta},\bm{S}]} \mid \bm{Q}_{[N]}^{[\bm{\Theta},\bm{S}]}, \bm{W}_{[M]}, \bm{W}_{M+1},\bm{\Theta}=M+1,\bm{S}=[M]\right)+ o(L) \\
&= L+H\left(\bm{A}_{[N]}^{[\bm{\Theta},\bm{S}]} \mid \bm{Q}_{[N]}^{[\bm{\Theta},\bm{S}]}, \bm{W}_{[M+1]},\bm{\Theta}=M+1,\bm{S}=[M]\right)+ o(L)  \label{ind} \\
&\geq L+H\left(\bm{A}_{\mathcal{T}}^{[\bm{\Theta},\bm{S}]}\mid \bm{Q}_{[N]}^{[\bm{\Theta},\bm{S}]}, \bm{W}_{[M+1]},\bm{\Theta}=M+1,\bm{S}=[M]\right)+ o(L)\\
&= L+H\left(\bm{A}_{\mathcal{T}}^{[\bm{\Theta},\bm{S}]}\mid \bm{Q}_{\mathcal{T}}^{[\bm{\Theta},\bm{S}]}, \bm{W}_{[M+1]},\bm{\Theta}=M+1,\bm{S}=[M]\right)+ o(L) \label{drop}\\
&\overset{(\ref{privacy})}{=} L+H\left(\bm{A}_{\mathcal{T}}^{[\bm{\Theta},\bm{S}]}\mid \bm{Q}_{\mathcal{T}}^{[\bm{\Theta},\bm{S}]}, \bm{W}_{[M+1]},\bm{\Theta}=M+2,\bm{S}=[M]\right)+ o(L)\\
&\ge L+H\left(\bm{A}_{\mathcal{T}}^{[\bm{\Theta},\bm{S}]}\mid\bm{Q}_{[N]}^{[\bm{\Theta},\bm{S}]}, \bm{W}_{[M+1]},\bm{\Theta}=M+2,\bm{S}=[M]\right)+ o(L) ,\label{eq:AT}
\end{align}
where (\ref{ind}) holds because messages are independent, and queries are independent of the messages. Equation (\ref{drop}) holds because given $\bm{\Theta}=M+1,\bm{S}=[M]$, $\bm{A}_{\mathcal{T}}^{[\bm{\Theta},\bm{S}]} \leftrightarrow \left(\bm{Q}_{\mathcal{T}}^{[\bm{\Theta},\bm{S}]}, \bm{W}_{[M+1]} \right)\leftrightarrow \bm{Q}_{[N]\backslash \mathcal{T}}^{[\bm{\Theta},\bm{S}]}$ is a Markov chain. There are a total of $\binom{N}{T}$  such subsets $\mathcal{T}$. Writing  \eqref{eq:AT} for all such subsets and adding those inequalities, we obtain,
\begin{align}
&\binom{N}{T}D(K,[M], M+1,M) \\
\geq &\binom{N}{T}L +\sum_{\mathcal{T}: \mathcal{T}\subset[N], |\mathcal{T}|=T}H\left(\bm{A}_{\mathcal{T}}^{[\bm{\Theta},\bm{S}]} \mid\bm{Q}_{[N]}^{[\bm{\Theta},\bm{S}]}, \bm{W}_{[M+1]},\bm{\Theta}=M+2,\bm{S}=[M]\right)+ o(L)\\
\geq &\binom{N}{T}L +\binom{N}{T}\frac{T}{N}H\left(\bm{A}_{[N]}^{[\bm{\Theta},\bm{S}]} \mid\bm{Q}_{[N]}^{[\bm{\Theta},\bm{S}]}, \bm{W}_{[M+1]},\bm{\Theta}=M+2,\bm{S}=[M]\right)+ o(L) \label{han}\\
= &\binom{N}{T}L +\binom{N}{T}\frac{T}{N}D(K,[M],M+2,M+1)+ o(L),
\end{align}
where (\ref{han}) follows from Han's inequality. Therefore,
\begin{align}
D(K,[M],M+1,M) \geq L + \frac{T}{N} D(K,[M],M+2,M+1) + o(L).
\end{align}
Proceeding along these lines, we have,
\begin{eqnarray}
D(K,[M],M+1,M) &\geq& L + \frac{T}{N} D(K,[M],M+2,M+1)+ o(L)\\
&\geq& L + \frac{T}{N} \left(L + \frac{T}{N} D(K,[M],M+3,M+2)\right)+ o(L)\\
&\geq& \cdots\\
&\geq& L + \frac{T}{N} \left(L + \cdots + \frac{T}{N}\left(L+\frac{T}{N} D(K,[M],K,K-1)\right)\right)+o(L)
\end{eqnarray}
where $D(K,[M],K,K-1) \geq L$. Therefore, 
\begin{eqnarray}
D(K,[M],M+1,M) &\geq& L + \frac{T}{N}L + \cdots + \left(\frac{T}{N}\right)^{K-M-1}L+ o(L)\\
&=& L\left(1+\frac{T}{N} + \cdots + \left(\frac{T}{N}\right)^{K-M-1}\right)+ o(L).
\end{eqnarray}
Noting that the above argument holds similarly for any $(\theta, S)$ gives us the upper bound on the rate of TPIR-PSI as
\begin{eqnarray}
R &=& \lim_{L\rightarrow\infty}\frac{L}{D} \\
&\leq& \left(1+\frac{T}{N}+\left(\frac{T}{N}\right)^2+\cdots+\left(\frac{T}{N}\right)^{K-M-1}\right)^{-1}.
\end{eqnarray}
Thus, the proof of converse for Theorem \ref{thm:1} is complete.

\section{Proof of Theorem \ref{thm:2}} \label{sec:T2}
\subsection{Achievability}
When $M=K-1$, the user can download the sum of all the messages from one database and get the desired message with side information. The rate is $1$, achieving the capacity. Note that in this case, common randomness among databases is not required. When $M<K-1$, the achievable scheme can directly use the scheme of STPIR \cite{Sun_Jafar_SPIR, Wang_Skoglund}, and the side information is simply not used. 

\subsection{Converse}
When $M=K-1$, it is obvious that $1$ is an upper bound, since any rates cannot be larger than $1$. When $M<K-1$, we show that $1-\frac{T}{N}$ is an upper bound. Let $\mathcal{S}$ be the set of subsets of $[K]$ of cardinality $M$, i.e., it contains all possible realizations of $\bm{S}$. We will need the following lemma.
\begin{lemma}\label{lemma:claim}
For all $S \in\mathcal{S}$, 
 $\theta, \theta' \in [K] \backslash S$, and $\mathcal{T} \subset [1:N], |\mathcal{T}| = T$,
\begin{align}
H\left(\bm{A}_{\mathcal{T}}^{[\bm{\Theta}, \bm{S}]} \mid \bm{Q}_{\mathcal{T}}^{[\bm{\Theta},\bm{S}]}, \bm{W}_{\bm{\Theta}},\bm{W}_{\bm{S}},\bm{\Theta}=\theta,\bm{S}=S\right)&=H\left(\bm{A}_{\mathcal{T}}^{[\bm{\Theta},\bm{S}]} \mid \bm{Q}_{\mathcal{T}}^{[\bm{\Theta},\bm{S}]}, \bm{W}_{\bm{\Theta}},\bm{W}_{\bm{S}},\bm{\Theta}=\theta', \bm{S}=S\right), \label{c1}\\
H\left(\bm{A}_{\mathcal{T}}^{[\bm{\Theta},\bm{S}]} \mid \bm{Q}_{\mathcal{T}}^{[\bm{\Theta},\bm{S}]}, \bm{W}_{\bm S},\bm{\Theta}=\theta,\bm{S}=S\right)&=H\left(\bm{A}_{\mathcal{T}}^{[\bm{\Theta},\bm{S}]} \mid \bm{Q}_{\mathcal{T}}^{[\bm{\Theta},\bm{S}]}, \bm{W}_{\bm S}, \bm{\Theta}=\theta', \bm{S}=S\right) \label{c2}. 
\end{align}
\end{lemma}

\proof  
It follows from the user-privacy constraint (\ref{privacy:symmetric}) and the non-negativity of mutual information, that for all $S \in\mathcal{S}$, $\mathcal{T} \subset [N], |\mathcal{T}| = T$ 
\begin{align}
I\left(\bm{\Theta} ; \bm{Q}_{\mathcal{T}}^{[\bm{\Theta},\bm{S}]}, \bm{A}_{\mathcal{T}}^{[\bm{\Theta},\bm{S}]}, \bm{W}_{[K]}\mid {\bm S}=S \right) = 0,
\end{align}
which implies that  $\forall \theta, \theta' \in [K] \backslash S$,
\begin{align}
H\left(\bm{Q}_{\mathcal{T}}^{[\bm{\Theta},\bm{S}]}, \bm{A}_{\mathcal{T}}^{[\bm{\Theta},\bm{S}]},\bm{W}_{\theta},\bm{W}_{\bm S} \mid \bm{\Theta}=\theta, \bm{S}=S\right)&=H\left(\bm{Q}_{\mathcal{T}}^{[\bm{\Theta},\bm{S}]}, \bm{A}_{\mathcal{T}}^{[\bm{\Theta},\bm{S}]},\bm{W}_{\theta},\bm{W}_{\bm{S}}\mid \bm{\Theta}=\theta', \bm{S}=S\right)\label{eq:l1}\\
H\left(\bm{Q}_{\mathcal{T}}^{[\bm{\Theta},\bm{S}]},\bm{W}_{\theta}, \bm{W}_{\bm S} \mid \bm{\Theta}=\theta, \bm{S}=S\right)&=H\left(\bm{Q}_{\mathcal{T}}^{[\bm{\Theta},\bm{S}]},\bm{W}_{\theta}, \bm{W}_{\bm S} \mid \bm{\Theta}=\theta', \bm{S}=S\right). \label{eq:l2}
\end{align}
Subtracting (\ref{eq:l2}) from (\ref{eq:l1}) yields \eqref{c1}.
Equation (\ref{c2}) is similarly obtained.
\hfill\QED

\paragraph{Proof of the bound $R \le 1 - T/N$.}
Let us start with an intuitive understanding of the upper bound, $R \le 1 - T/N$. Due to  database privacy, given the side information, the answers from all  $N$ databases should be independent of the non-queried messages. At the same time, the answers from any  $T$ databases should contain no information about the queried message index since the user privacy must be preserved. Combining these two facts, given the side information, the answers from any  $T$ databases should contain no information about \emph{any} individual message, whether desired or undesired. As a result, the useful information about the desired message must come  from the remaining $N-T$ databases. Thus, the download per database must be at least $1/(N-T)$ times the entropy of the desired message. 

The formal proof is as follows. Since $M<K-1$, for any $S\in\mathcal{S}$, there exist at least $2$ messages that are not in the set $S$.  Any feasible STPIR-PSI scheme  must satisfy the database-privacy constraint (\ref{noleak}), 
\begin{align}
0&=I\left(\bm{W}_{\overline{(\bm{\Theta,S})}}; \bm{Q}_{[N]}^{[\bm{\Theta},\bm{S}]}, \bm{A}_{[N]}^{[\bm{\Theta},\bm{S}]}   \mid \bm{W}_{\bm S}, \bm{S},\bm{\Theta}\right) \label{eq:76}
\end{align}
Therefore, $\forall \mathcal{T} \subset [1:N], |{\mathcal{T}}|=T, \forall S \in \mathcal{S},$ and for all distinct $\theta,\theta' \in[K] \backslash S$,
\begin{align}
0&= I\left(\bm{W}_{\theta'}; \bm{A}_{\mathcal{T}}^{[\bm{\Theta},\bm{S}]},\bm{Q}_{\mathcal{T}}^{[\bm{\Theta},\bm{S}]}  \mid  \bm{W}_{\bm S},\bm{\Theta}=\theta,\bm{S}=S\right) \label{eq:same1}\\
&= H\left(\bm{A}_{\mathcal{T}}^{[\bm{\Theta},\bm{S}]} \mid \bm{Q}_{\mathcal{T}}^{[\bm{\Theta},\bm{S}]}, \bm{W}_{\bm S},\bm{\Theta}=\theta, \bm{S}=S\right) - H\left(\bm{A}_{\mathcal{T}}^{[\bm{\Theta},\bm{S}]} \mid  \bm{Q}_{\mathcal{T}}^{[\bm{\Theta},\bm{S}]}, \bm{W}_{\bm S}, \bm{W}_{\theta'},\bm{\Theta}=\theta, \bm{S}=S\right) \\
&\overset{(\ref{c1})}{=} H\left(\bm{A}_{\mathcal{T}}^{[\bm{\Theta},\bm{S}]} \mid \bm{Q}_{\mathcal{T}}^{[\bm{\Theta},\bm{S}]}, \bm{W}_{\bm S},\bm{\Theta}=\theta, \bm{S}=S\right) - H\left(\bm{A}_{\mathcal{T}}^{[\bm{\Theta},\bm{S}]} \mid  \bm{Q}_{\mathcal{T}}^{[\bm{\Theta},\bm{S}]}, \bm{W}_{\bm S}, \bm{W}_{\theta'},\bm{\Theta}=\theta', \bm{S}=S\right), \label{eq:same}
\end{align}
where (\ref{eq:same1}) is based on \eqref{eq:76}. According to the correctness condition,
\begin{align}
L &= H\left(\bm{W}_{\theta'}\right) \overset{(\ref{correct:PSI})}{=} I\left(\bm{W}_{\theta'}; \bm{A}_{[N]}^{[\bm{\Theta},\bm{S}]} \mid  \bm{W}_{\bm{S}}, \bm{Q}_{[N]}^{[\bm{\Theta},\bm{S}]},\bm{\Theta}=\theta', \bm{S}=S\right) + o(L)\\
&= H\left(\bm{A}_{[N]}^{[\bm{\Theta},\bm{S}]} \mid  \bm{W}_{\bm S} , \bm{Q}_{[N]}^{[\bm{\Theta},\bm{S}]},\bm{\Theta}=\theta', \bm{S}=S\right) -H\left(\bm{A}_{[N]}^{[\bm{\Theta},\bm{S}]} \mid \bm{W}_{\theta'}, \bm{W}_{\bm S} , \bm{Q}_{[N]}^{[\bm{\Theta},\bm{S}]},\bm{\Theta}=\theta', \bm{S}=S\right) + o(L) \\
&\leq H\left(\bm{A}_{[N]}^{[\bm{\Theta},\bm{S}]} \mid  \bm{W}_{\bm S} , \bm{Q}_{[N]}^{[\bm{\Theta},\bm{S}]},\bm{\Theta}=\theta', \bm{S}=S\right) -H\left(\bm{A}_{\mathcal{T}}^{[\bm{\Theta},\bm{S}]} \mid \bm{W}_{\theta'}, \bm{W}_{\bm S} , \bm{Q}_{[N]}^{[\bm{\Theta},\bm{S}]},\bm{\Theta}=\theta', \bm{S}=S\right) + o(L) \\
&= H\left(\bm{A}_{[N]}^{[\bm{\Theta},\bm{S}]} \mid  \bm{W}_{\bm S} , \bm{Q}_{[N]}^{[\bm{\Theta},\bm{S}]},\bm{\Theta}=\theta', \bm{S}=S\right) -H\left(\bm{A}_{\mathcal{T}}^{[\bm{\Theta},\bm{S}]} \mid \bm{W}_{\theta'}, \bm{W}_{\bm S} , \bm{Q}^{[\bm{\Theta},\bm{S}]}_{\mathcal{T}},\bm{\Theta}=\theta', \bm{S}=S\right) + o(L),\label{eq:dep} \\
&\overset{(\ref{eq:same})}{=} H\left(\bm{A}_{[N]}^{[\bm{\Theta},\bm{S}]} \mid  \bm{W}_{\bm S} , \bm{Q}_{[N]}^{[\bm{\Theta},\bm{S}]},\bm{\Theta}=\theta', \bm{S}=S\right) -H\left(\bm{A}_{\mathcal{T}}^{[\bm{\Theta},\bm{S}]} \mid  \bm{W}_{\bm S} , \bm{Q}^{[\bm{\Theta},\bm{S}]}_{\mathcal{T}},\bm{\Theta}=\theta, \bm{S}=S\right) + o(L) \\
&\overset{(\ref{c2})}{=} H\left(\bm{A}_{[N]}^{[\bm{\Theta},\bm{S}]} \mid  \bm{W}_{\bm S} , \bm{Q}_{[N]}^{[\bm{\Theta},\bm{S}]},\bm{\Theta}=\theta', \bm{S}=S\right) -H\left(\bm{A}_{\mathcal{T}}^{[\bm{\Theta},\bm{S}]} \mid  \bm{W}_{\bm S} , \bm{Q}^{[\bm{\Theta},\bm{S}]}_{\mathcal{T}},\bm{\Theta}=\theta', \bm{S}=S\right) + o(L)\\
&\leq H\left(\bm{A}_{[N]}^{[\bm{\Theta},\bm{S}]} \mid  \bm{W}_{\bm S} , \bm{Q}_{[N]}^{[\bm{\Theta},\bm{S}]},\bm{\Theta}=\theta', \bm{S}=S\right) -H\left(\bm{A}_{\mathcal{T}}^{[\bm{\Theta},\bm{S}]} \mid  \bm{W}_{\bm S} , \bm{Q}_{[N]}^{[\bm{\Theta},\bm{S}]},\bm{\Theta}=\theta', \bm{S}=S\right) + o(L),\label{eq:last}
\end{align} 
where \eqref{eq:dep} follows since given $\bm{\Theta}=\theta',\bm{S}=S$, $\bm{A}_{\mathcal{T}}^{[\bm{\Theta},\bm{S}]} \leftrightarrow \left(\bm{Q}_{\mathcal{T}}^{[\bm{\Theta},\bm{S}]}, \bm{W}_{\bm{S}} \right)\leftrightarrow \bm{Q}_{[N] \backslash \mathcal{T}}^{[\bm{\Theta},\bm{S}]}$ is a Markov chain.
Writing (\ref{eq:last}) for all $\mathcal{T} \subset [1:N], |\mathcal{T}|=T$,  and adding  those inequalities we obtain,
\begin{align}
\binom{N}{T} L &\leq \binom{N}{T} H\left(\bm{A}_{[N]}^{[\bm{\Theta},\bm{S}]} \mid  \bm{W}_{\bm S} , \bm{Q}_{[N]}^{[\bm{\Theta},\bm{S}]},\bm{\Theta}=\theta', \bm{S}=S\right)\notag\\
& \hspace{2.5cm}- \sum_{\mathcal{T}: |\mathcal{T}|=T} H\left(\bm{A}_{\mathcal{T}}^{[\bm{\Theta},\bm{S}]} \mid  \bm{W}_{\bm S} , \bm{Q}_{[N]}^{[\bm{\Theta},\bm{S}]},\bm{\Theta}=\theta', \bm{S}=S\right) + o(L) \\
&\leq \binom{N}{T} H\left(\bm{A}_{[N]}^{[\bm{\Theta},\bm{S}]} \mid  \bm{W}_{\bm S} , \bm{Q}_{[N]}^{[\bm{\Theta},\bm{S}]},\bm{\Theta}=\theta', \bm{S}=S\right)\notag\\
& \hspace{2.5cm}-  \binom{N}{T} \frac{T}{N} H\left(\bm{A}_{[N]}^{[\bm{\Theta},\bm{S}]} \mid  \bm{W}_{\bm S} , \bm{Q}_{[N]}^{[\bm{\Theta},\bm{S}]},\bm{\Theta}=\theta', \bm{S}=S\right) + o(L)\label{han2}  \\
&= \binom{N}{T} \left(1-\frac{T}{N}\right)H\left(\bm{A}_{[N]}^{[\bm{\Theta},\bm{S}]} \mid  \bm{W}_{\bm S} , \bm{Q}_{[N]}^{[\bm{\Theta},\bm{S}]},\bm{\Theta}=\theta', \bm{S}=S\right)+ o(L), \label{eq:ee}
\end{align}
where (\ref{han2}) is due to Han's inequality.
Since this inequality is true for all $S\in\mathcal{S}, \theta'\in[K]\backslash S$, it is also true when averaged across them, so,
\begin{align}
\binom{N}{T} L &\leq  \binom{N}{T} \left(1-\frac{T}{N}\right)H\left(\bm{A}_{[N]}^{[\bm{\Theta},\bm{S}]} \mid  \bm{W}_{\bm S} , \bm{Q}_{[N]}^{[\bm{\Theta},\bm{S}]},\bm{\Theta}, \bm{S}\right)+ o(L) \\
&\leq  \binom{N}{T} \left(1-\frac{T}{N}\right)H\left(\bm{A}_{[N]}^{[\bm{\Theta},\bm{S}]}\right)+ o(L)  \label{drop2}\\
&\leq \binom{N}{T} \left(1-\frac{T}{N}\right) D + o(L),
\label{eq:asame} 
\end{align}
where (\ref{drop2}) holds because dropping conditioning does not reduce entropy. Therefore, $R = \lim_{L\rightarrow \infty}\frac{L}{D} \leq 1 - \frac{T}{N}$, and we have shown that the rate of any feasible STPIR-SI scheme cannot be more than $1- \frac{T}{N}$.

\paragraph{Proof of the bound $\rho \geq T/(N-T)$.}
Let us first explain the intuition behind this bound on the size of the common randomness $U$ that should be available to all databases but not to the user. We have already shown that the normalized size of the answer from any individual database must be at least $L/(N-T)$. Due to the user and database privacy constraints, the answers from any  $T$ databases are independent of the messages. Therefore, to ensure database privacy, the amount of common randomness must be no smaller than the size of the answers from $T$ databases.

The formal proof is as follows. Suppose a feasible STPIR-PSI scheme exists that achieves a non-zero rate. Then we show that it must satisfy $\rho \geq T/(N-T)$. For $\bm{S}=S \in \mathcal{S}$ and for $\Theta=\theta \in [K]\backslash S$, consider the answering strings $\bm{A}_1^{[\bm{\Theta},\bm{S}]}, \cdots, \bm{A}_N^{[\bm{\Theta},\bm{S}]}$ and the side information $\bm{W}_{\bm{S}}$, from which the user can retrieve $\bm{W}_{\theta}$. According to the database-privacy constraint, we have
\begin{align}
0 &= I\left(\bm{W}_{\overline{(\theta,S)}}~; \bm{A}_{[N]}^{[\bm{\Theta},\bm{S}]}   \mid   \bm{W}_{\bm{S}}, \bm{Q}_{[N]}^{[\bm{\Theta},\bm{S}]}, \bm{\Theta}=\theta,\bm{S}=S\right) \\
&\overset{(\ref{correct:PSI})}{=} I\left(\bm{W}_{\overline{(\theta,S)}}~; \bm{A}_{[N]}^{[\bm{\Theta},\bm{S}]}  , \bm{W}_\theta \mid   \bm{W}_{\bm{S}}, \bm{Q}_{[N]}^{[\bm{\Theta},\bm{S}]}, \bm{\Theta}=\theta,\bm{S}=S\right) + o(L)\\
&\overset{(\ref{indep})}{=} I\left(\bm{W}_{\overline{(\theta,S)}}~; \bm{A}_{[N]}^{[\bm{\Theta},\bm{S}]}   \mid   \bm{W}_{\theta}, \bm{W}_{\bm S},  \bm{Q}_{[N]}^{[\bm{\Theta},\bm{S}]}, \bm{\Theta}=\theta,\bm{S}=S\right) + o(L) \label{eq:e2} \\
&\geq I\left(\bm{W}_{\overline{(\theta,S)}}~; \bm{A}_{\mathcal{T}}^{[\bm{\Theta},\bm{S}]}   \mid   \bm{W}_{\theta}, \bm{W}_{\bm S},  \bm{Q}_{[N]}^{[\bm{\Theta},\bm{S}]}, \bm{\Theta}=\theta,\bm{S}=S\right) + o(L)\\
&= H\left(\bm{A}_{\mathcal{T}}^{[\bm{\Theta},\bm{S}]}   \mid   \bm{W}_{\theta}, \bm{W}_{\bm S},  \bm{Q}_{[N]}^{[\bm{\Theta},\bm{S}]}, \bm{\Theta}=\theta,\bm{S}=S\right) -H\left(\bm{A}_{\mathcal{T}}^{[\bm{\Theta},\bm{S}]}   \mid   \bm{W}_{[K]},  \bm{Q}_{[N]}^{[\bm{\Theta},\bm{S}]}, \bm{\Theta}=\theta,\bm{S}=S\right)+ o(L)\\
&\overset{(\ref{ansdet})}{=} H\left(\bm{A}_{\mathcal{T}}^{[\bm{\Theta},\bm{S}]}   \mid   \bm{W}_{\theta}, \bm{W}_{\bm S},  \bm{Q}_{[N]}^{[\bm{\Theta},\bm{S}]}, \bm{\Theta}=\theta,\bm{S}=S\right) -H\left(\bm{A}_{\mathcal{T}}^{[\bm{\Theta},\bm{S}]}   \mid   \bm{W}_{[K]},  \bm{Q}_{[N]}^{[\bm{\Theta},\bm{S}]}, \bm{\Theta}=\theta,\bm{S}=S\right) \notag \\
&\hspace{1cm}+ H\left(\bm{A}_{\mathcal{T}}^{[\bm{\Theta},\bm{S}]}   \mid   \bm{W}_{[K]},  \bm{Q}_{[N]}^{[\bm{\Theta},\bm{S}]},\bm{U}, \bm{\Theta}=\theta,\bm{S}=S\right) \label{eq:e3}  + o(L)\\
&= H\left(\bm{A}_{\mathcal{T}}^{[\bm{\Theta},\bm{S}]}   \mid   \bm{W}_{\theta}, \bm{W}_{\bm S},  \bm{Q}_{[N]}^{[\bm{\Theta},\bm{S}]}, \bm{\Theta}=\theta,\bm{S}=S\right)  -I\left(\bm{U};\bm{A}_{\mathcal{T}}^{[\bm{\Theta},\bm{S}]}   \mid   \bm{W}_{[K]},  \bm{Q}_{[N]}^{[\bm{\Theta},\bm{S}]}, \bm{\Theta}=\theta,\bm{S}=S\right) + o(L)\\
&\geq H\left(\bm{A}_{\mathcal{T}}^{[\bm{\Theta},\bm{S}]}   \mid   \bm{W}_{\theta}, \bm{W}_{\bm S},  \bm{Q}^{[\bm{\Theta},\bm{S}]}_{\mathcal{T}}, \bm{\Theta}=\theta,\bm{S}=S\right)  -H(\bm{U}) + o(L)\\
&\overset{(\ref{eq:same})}{=} H\left(\bm{A}_{\mathcal{T}}^{[\bm{\Theta},\bm{S}]}   \mid   \bm{W}_{\bm S},  \bm{Q}^{[\bm{\Theta},\bm{S}]}_{\mathcal{T}}, \bm{\Theta}=\theta',\bm{S}=S\right)  -H(\bm{U}) + o(L)\\
&\overset{(\ref{c2})}{=} H\left(\bm{A}_{\mathcal{T}}^{[\bm{\Theta},\bm{S}]}   \mid   \bm{W}_{\bm S},  \bm{Q}^{[\bm{\Theta},\bm{S}]}_{\mathcal{T}}, \bm{\Theta}=\theta,\bm{S}=S\right)  -H(\bm{U}) + o(L).\label{eq:e6}
\end{align}
Adding (\ref{eq:e6}) for  all $\mathcal{T} \subset [N], |\mathcal{T}|=T$, we obtain,
\begin{align}
0 &\geq \sum_{\mathcal{T}:~\mathcal{T}\subset[N], |\mathcal{T}|=T} H\left(\bm{A}_{\mathcal{T}}^{[\bm{\Theta},\bm{S}]}   \mid   \bm{W}_{\bm S},  \bm{Q}^{[\bm{\Theta},\bm{S}]}_{\mathcal{T}}, \bm{\Theta}=\theta,\bm{S}=S\right) - \binom{N}{T}  H(\bm{U}) + o(L)\\
&\geq \binom{N}{T} \frac{T}{N} H\left(\bm{A}_{[N]}^{[\bm{\Theta},\bm{S}]}   \mid   \bm{W}_{\bm S},  \bm{Q}^{[\bm{\Theta},\bm{S}]}_{\mathcal{T}}, \bm{\Theta}=\theta,\bm{S}=S\right) - \binom{N}{T}  H(\bm{U}) + o(L)\label{han3}\\
&\geq \binom{N}{T} \frac{T}{N} H\left(\bm{A}_{[N]}^{[\bm{\Theta},\bm{S}]}   \mid   \bm{W}_{\bm S},  \bm{Q}_{[N]}^{[\bm{\Theta},\bm{S}]}, \bm{\Theta}=\theta,\bm{S}=S\right) - \binom{N}{T}  H(\bm{U}) + o(L)\\
&\overset{(\ref{eq:ee})}{\geq} \binom{N}{T} \frac{T}{N}\left( \frac{N}{N-T}\right) L - \binom{N}{T}  H(\bm{U}) \label{eq:e7} + o(L).\\
\Rightarrow H(\bm{U}) &\geq \left(\frac{T}{N-T}\right) L + o(L),\\
\Rightarrow \rho &= \frac{H(\bm{U})}{ L} \geq \frac{T}{N-T} \label{eq:rhobound} ~~~\mbox{(letting $L \rightarrow \infty$)}.
\end{align} 
Note that (\ref{han3}) is due to Han's inequality. Thus the amount of common randomness normalized by the message size for any feasible STPIR-PSI scheme cannot be less than $T/(N-T)$.


\section{Conclusion}\label{sec:conclusion}
The capacity of TPIR-PSI and the capacity of STPIR-PSI  are characterized. As a special case of TPIR-PSI obtained by setting  $T=1$, the result solves the capacity of PIR-PSI, an open problem highlighted by Kadhe et al. in \cite{Kadhe_Garcia_Heidarzadeh_Rouayheb_Sprintson}. Notably, the results of our work (initially limited to capacity of PIR-PSI for $T=1$ as  reported in our original ArXiv posting in 2017 \cite{Chen_Wang_Jafar_v1}) have subsequently been generalized to multi-message PIR-PSI in \cite{Shariatpanahi_Siavoshani_Maddah}. Other generalizations, e.g.,  PIR-PSI with multi-round communication, secure and/or coded storage, remain promising directions for future work, as is the capacity characterization for PIR-NSI which remains open for multi-server settings.

\bibliographystyle{IEEEtran}
\bibliography{Thesis}

\begin{thebibliography}{10}
\providecommand{\url}[1]{#1}
\csname url@samestyle\endcsname
\providecommand{\newblock}{\relax}
\providecommand{\bibinfo}[2]{#2}
\providecommand{\BIBentrySTDinterwordspacing}{\spaceskip=0pt\relax}
\providecommand{\BIBentryALTinterwordstretchfactor}{4}
\providecommand{\BIBentryALTinterwordspacing}{\spaceskip=\fontdimen2\font plus
\BIBentryALTinterwordstretchfactor\fontdimen3\font minus
  \fontdimen4\font\relax}
\providecommand{\BIBforeignlanguage}[2]{{%
\expandafter\ifx\csname l@#1\endcsname\relax
\typeout{** WARNING: IEEEtran.bst: No hyphenation pattern has been}%
\typeout{** loaded for the language `#1'. Using the pattern for}%
\typeout{** the default language instead.}%
\else
\language=\csname l@#1\endcsname
\fi
#2}}
\providecommand{\BIBdecl}{\relax}
\BIBdecl

\bibitem{PIRfirstjournal}
B.~Chor, E.~Kushilevitz, O.~Goldreich, and M.~Sudan, ``{Private Information
  Retrieval},'' \emph{Journal of the ACM (JACM)}, vol.~45, no.~6, pp. 965--981,
  1998.

\bibitem{Yekhanin}
S.~Yekhanin, ``{Private Information Retrieval},'' \emph{Communications of the
  ACM}, vol.~53, no.~4, pp. 68--73, 2010.

\bibitem{Beimel_Ishai_Kushilevitz}
A.~Beimel, Y.~Ishai, and E.~Kushilevitz, ``{General constructions for
  information-theoretic private information retrieval},'' \emph{Journal of
  Computer and System Sciences}, vol.~71, no.~2, pp. 213--247, 2005.

\bibitem{William}
W.~Gasarch, ``{A Survey on Private Information Retrieval},'' in \emph{Bulletin
  of the EATCS}, 2004.

\bibitem{CPIR}
R.~Ostrovsky and W.~E. Skeith~III, ``{A Survey of Single-database Private
  Information Retrieval: Techniques and Applications},'' in \emph{Public Key
  Cryptography--PKC 2007}.\hskip 1em plus 0.5em minus 0.4em\relax Springer,
  2007, pp. 393--411.

\bibitem{SymPIR}
Y.~Gertner, Y.~Ishai, E.~Kushilevitz, and T.~Malkin, ``Protecting data privacy
  in private information retrieval schemes,'' in \emph{Proceedings of the
  thirtieth annual ACM symposium on Theory of computing}.\hskip 1em plus 0.5em
  minus 0.4em\relax ACM, 1998, pp. 151--160.

\bibitem{Shah_Rashmi_Kannan}
N.~Shah, K.~Rashmi, and K.~Ramchandran, ``{One Extra Bit of Download Ensures
  Perfectly Private Information Retrieval},'' in \emph{Proceedings of IEEE
  International Symposium on Information Theory (ISIT)}, 2014, pp. 856--860.

\bibitem{Sun_Jafar_BIAPIR}
H.~Sun and S.~A. Jafar, ``Blind interference alignment for private information
  retrieval,'' \emph{IEEE International Symposium on Information Theory
  (ISIT)}, pp. 560--564, 2016.

\bibitem{Sun_Jafar_PIR}
------, ``{The Capacity of Private Information Retrieval},'' \emph{IEEE
  Transactions on Information Theory}, vol.~63, no.~7, pp. 4075--4088, July
  2017.

\bibitem{Tajeddine_Rouayheb}
R.~Tajeddine, O.~W. Gnilke, and S.~El~Rouayheb, ``{Private Information
  Retrieval from {MDS} Coded Data in Distributed Storage Systems},'' \emph{IEEE
  Transactions on Information Theory}, 2018.

\bibitem{Chan_Ho_Yamamoto}
T.~H. Chan, S.-W. Ho, and H.~Yamamoto, ``{Private Information Retrieval for
  Coded Storage},'' \emph{Proceedings of IEEE International Symposium on
  Information Theory (ISIT)}, pp. 2842--2846, 2015.

\bibitem{Banawan_Ulukus}
K.~Banawan and S.~Ulukus, ``{The Capacity of Private Information Retrieval from
  Coded Databases},'' \emph{IEEE Transactions on Information Theory}, vol.~64,
  no.~3, pp. 1945--1956, 2018.

\bibitem{Sun_Jafar_TPIR}
H.~Sun and S.~A. Jafar, ``{The Capacity of Robust Private Information Retrieval
  with Colluding Databases},'' \emph{IEEE Transactions on Information Theory},
  vol.~64, no.~4, pp. 2361--2370, 2018.

\bibitem{Sun_Jafar_SPIR}
------, ``The capacity of symmetric private information retrieval,'' \emph{IEEE
  Transactions on Information Theory}, vol.~65, no.~1, pp. 322--329, 2019.

\bibitem{Wang_Skoglund}
Q.~Wang and M.~Skoglund, ``{Symmetric Private Information Retrieval For MDS
  Coded Distributed Storage},'' \emph{2017 IEEE International Conference on
  Communications (ICC)}, pp. 1--6, 2017.

\bibitem{Banawan_Ulukus_MPIR}
K.~Banawan and S.~Ulukus, ``Multi-message private information retrieval:
  Capacity results and near-optimal schemes,'' \emph{arXiv preprint
  arXiv:1702.01739}, 2017.

\bibitem{Sun_Jafar_MPIR}
H.~Sun and S.~A. Jafar, ``{Multiround Private Information Retrieval: Capacity
  and Storage Overhead},'' \emph{IEEE Transactions on Information Theory},
  vol.~64, no.~8, pp. 5743--5754, August 2018.

\bibitem{Jia_Sun_Jafar_XSTPIR}
Z.~Jia, H.~Sun, and S.~A. Jafar, ``{Cross Subspace Alignment and the Asymptotic
  Capacity of $X$-Secure $T$-Private Information Retrieval},'' \emph{arXiv
  preprint arXiv:1808.07457}, 2018.

\bibitem{Tandon_CPIR}
R.~Tandon, ``The capacity of cache aided private information retrieval,''
  \emph{55th Annual Allerton Conference on Communication, Control, and
  Computing (Allerton)}, 2017.

\bibitem{Kadhe_Garcia_Heidarzadeh_Rouayheb_Sprintson}
S.~Kadhe, B.~Garcia, A.~Heidarzadeh, S.~E. Rouayheb, and A.~Sprintson,
  ``Private information retrieval with side information: The single server
  case,'' \emph{55th Annual Allerton Conference on Communication, Control, and
  Computing (Allerton)}, 2017.

\bibitem{Wei_Banawan_Ulukus}
Y.-P. Wei, K.~Banawan, and S.~Ulukus, ``Fundamental limits of cache-aided
  private information retrieval with unknown and uncoded prefetching,''
  \emph{IEEE Transactions on Information Theory}, vol.~65, no.~5, pp.
  3215--3232, 2019.

\bibitem{Wei_Banawan_Ulukus_Side}
------, ``Cache-aided private information retrieval with partially known
  uncoded prefetching: fundamental limits,'' \emph{IEEE Jour. on Selected Areas
  in Communications}, vol.~36, no.~6, pp. 1126--1139, 2018.

\bibitem{Rouayheb_Sprintson_single_coded}
A.~Heidarzadeh, F.~Kazemi, and A.~Sprintson, ``Capacity of single-server
  single-message private information retrieval with coded side information,''
  \emph{arXiv preprint arXiv:1806.00661}, 2018.

\bibitem{Rouayheb_Sprintson_single_multi}
A.~Heidarzadeh, B.~Garcia, S.~Kadhe, S.~E. Rouayheb, and A.~Sprintson, ``On the
  capacity of single-server multi-message private information retrieval with
  side information,'' \emph{arXiv preprint arXiv:1807.09908}, 2018.

\bibitem{Rouayheb_Sprintson_single_multi_individual}
A.~Heidarzadeh, S.~Kadhe, S.~E. Rouayheb, and A.~Sprintson, ``Single-server
  multi-message individually-private information retrieval with side
  information,'' \emph{arXiv preprint arXiv:1901.07509}, 2019.

\bibitem{Rouayheb_Sprintson_single_single_online}
F.~Kazemi, E.~Karimi, A.~Heidarzadeh, and A.~Sprintson, ``Single-server
  single-message online private information retrieval with side information,''
  \emph{arXiv preprint arXiv:1901.07748}, 2019.

\bibitem{Li_Gastpar}
S.~Li and M.~Gastpar, ``Converse for multi-server single-message pir with side
  information,'' \emph{arXiv preprint arXiv:1809.09861}, 2018.

\bibitem{Li_Gastpar_muPIR}
------, ``Single-server multi-user private information retrieval with side
  information,'' \emph{Proceedings of IEEE International Symposium on
  Information Theory (ISIT)}, 2018.

\bibitem{Chen_Wang_Jafar_v1}
Z.~Chen, Z.~Wang, and S.~Jafar, ``The capacity of private information retrieval
  with private side information,'' \emph{arXiv preprint arXiv:1709.03022v1},
  Sep. 2017.

\bibitem{Shariatpanahi_Siavoshani_Maddah}
S.~P. Shariatpanahi, M.~J. Siavoshani, and M.~A. Maddah-Ali, ``Multi-message
  private information retrieval with private side information,'' \emph{arXiv
  preprint arXiv:1805.11892}, 2018.

\end{thebibliography}

\end{document}